\documentstyle[twoside,fleqn,espcrc2]{article}

\input epsf

\newcommand{\AmS}{{\protect\the\textfont2
  A\kern-.1667em\lower.5ex\hbox{M}\kern-.125emS}}

\title{Grassmann integrals by machine}

\author{Michael Creutz
\address{Physics Department, Brookhaven
 National Laboratory, PO Box 5000, Upton, NY 11973-5000, USA\\
 creutz@bnl.gov
}
\thanks{This manuscript has been authored under contract number
DE-AC02-98CH10886 with the U.S.~Department of Energy.  Accordingly,
the U.S. Government retains a non-exclusive, royalty-free license to
publish or reproduce the published form of this contribution, or allow
others to do so, for U.S.~Government purposes.}
}
 
\begin{document}
 
\begin{abstract}
I present a numerical algorithm for direct evaluation of multiple
Grassmann integrals.  The approach is exact and suffers no Fermion
sign problems.  Memory requirements grow exponentially with the
interaction range and the transverse size of the system.  Low
dimensional systems of order a thousand Grassmann variables can be
evaluated on a workstation.
\end{abstract}
 
\maketitle

In quantum field theory fermions are usually treated via integrals
over anti-commuting Grassmann variables \cite{grassmann}, providing an
elegant framework for the formal establishment of Feynman perturbation
theory.  With non-perturbative approaches, such as Monte Carlo studies
on the lattice, these objects are more problematic.  Essentially all
approaches formally integrate the fermionic fields in terms of
determinants depending only on bosonic fields.  When a background
fermion density is present, as for baryon rich regions of heavy ion
scattering, these determinants are not positive, making Monte Carlo
evaluations tedious on any but the smallest systems\cite{barbour}.
This problem also appears in studies of many electron systems doped
away from half filling.

Here I explore the possibility of directly evaluating the fermionic
integrals, doing the necessary combinatorics on a
computer\cite{letter}. This is inevitably a rather tedious task, with
the expected effort growing exponentially with volume.  Nevertheless,
in the presence of the sign problem, all other known algorithms are
also exponential.  My main result is that this growth can be
controlled to a transverse section of the system.  I illustrate the
technique with a low dimensional system involving of order a thousand
Grassmann variables.

I begin with a set of anti-commuting Grassmann variables $\{\psi\}$,
satisfying $\psi_i\psi_j+\psi_j\psi_i=0$. Integration is uniquely
determined up to an overall normalization by requiring linearity and
``translation'' invariance {$\int d\psi f(\psi+\psi^\prime) = \int
d\psi f(\psi)$}.  I normalize things so that
\begin{equation}
\int d\psi\ \psi = 1 \qquad \int d\psi\ 1 = 0.
\label{basic}
\end{equation}
Consider an arbitrary action $S(\psi)$ inserted into a path integral.
I want to evaluate
\begin{equation}
Z=\int d\psi_n \ldots d\psi_1\ e^{S(\psi)}.
\label{pathintegral}
\end{equation}
Formally this requires expanding the exponent and keeping all terms
containing exactly one factor of each $\psi_i$.

I first reduce the required expansion into operator manipulations in a
Fock space.  Introduce a fermionic creation-annihilation pair for each
fermionic field, $\psi_i \leftrightarrow \{a_i^\dagger,a_i\}$.  These
satisfy the usual relations $ [a_i,a_j^\dagger]_+=\delta_{ij}.  $ The
space is built by applying creation operators to the vacuum, which
satisfies $a_i{|0\rangle}=0$.  It is convenient to introduce the
completely occupied ``full'' state $ {|F\rangle} \equiv
a_n^\dagger...a_2^\dagger a_1^\dagger {|0\rangle}.  $ Then I rewrite
my basic path integral as the matrix element
\begin{equation}
Z\ =\ {\langle 0|}\  e^{S(a)}\  {|F\rangle}.
\label{me1}
\end{equation}
Expanding {$e^S$}, a non-vanishing contribution requires one factor of
{$a_i$} for each Fermion.  This is the same rule as for Grassmann
integration.

I now manipulate this expression towards a sequential evaluation.
Select a single variable $\psi_i$ and define $S_i(a)$ as all terms
from the action involving a factor of $a_i$.  I define the complement
$\tilde S_i$ as anything else, so that $S=S_i+\tilde S_i$.  I assume a
bosonic action so that {$S_i$} and {$\tilde S_i$} commute and $
Z={\langle 0 |} e^{\tilde S_i} e^{S_i}{|F\rangle}.  $ Since $\tilde
S_i$ contains no factors of $a_i$, the occupation number for that
variable, $n_i=a_i^\dagger a_i$, vanishes between the two factors.  I
thus can insert a projection operator $1-n_i$
\begin{equation}
Z={\langle 0 |} e^{\tilde S_i}\ (1-n_i)\ e^{S_i}{|F\rangle}.
\label{me3}
\end{equation}
Since $1-n_i$ projects out an empty state at location $i$, I trivially
have $a_i\ (1-n_i)=0$.  I can replace {$\tilde S_i$} with the full
action.  Also, since $a_i^2=0$, the right hand factor expands as
$e^{S_i}= 1+S_i$, giving
\begin{equation}
Z={\langle 0 |} e^{S}\ (1-n_i)\ (1+S_i){|F\rangle}.
\label{me5}
\end{equation}
Repetition gives my main result
\begin{equation}
Z=
\left\langle 0 \left\vert
 \prod_i \left( (1-n_i) (1+S_i)\right )
\right\vert F\right\rangle.
\label{central}
\end{equation}
assuming constants in the action are removed.

Eq.~(\ref{central}) summarizes the basic procedure.  Create an
associative array (hash table) to store general states of the Fock
space.  For a given state $ \vert \psi \rangle=\sum_s \chi_s
|s\rangle$ store the numbers $\chi_s$ labeled by $|s\rangle$.
Initially this table only contains the one entry for the full state.
The algorithm then loops over the Grassmann variables.  For a given
$\psi_i$, first apply $(1+S_i)$ to the stored state.  Then empty the
location with the projector $1-n_i$.  After all sites are integrated
over, only the empty state survives, with the desired integral as its
coefficient.

\begin{figure}
\epsfxsize .8\hsize
\centerline {\epsfbox{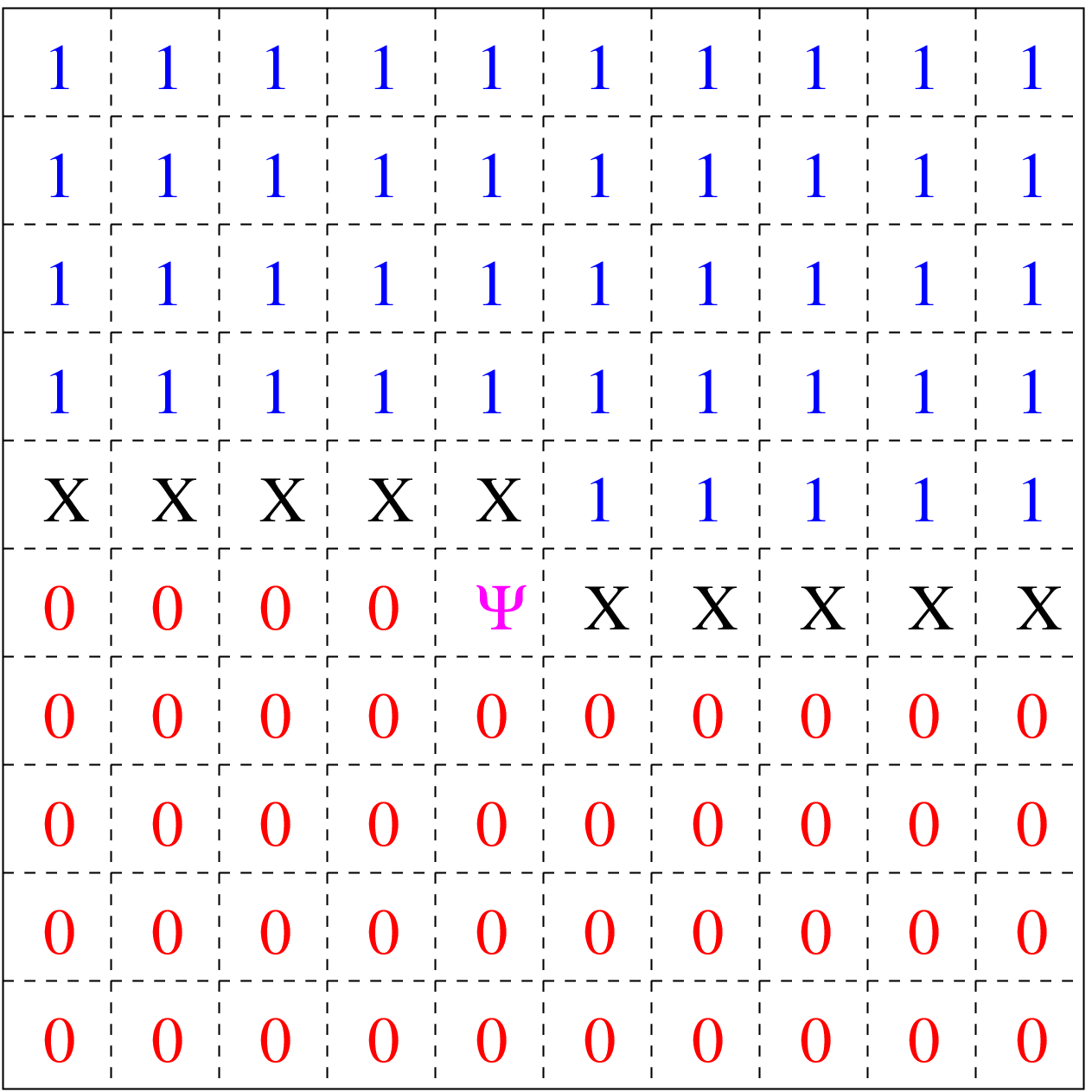}}
\vskip -.2in
\caption {Integrating out sequentially, visited sites are empty and
out of range sites are filled.  When integrating the site labeled
$\psi$, only those sites labeled ``X'' are undetermined.}
\label{latfig}
\end{figure}

The advantages appear with a local interaction.  All sites previously
visited are empty, and involve no information.  Unvisited locations
outside the interaction range are still filled, and also involve no
storage.  All relevant states are nontrivial only for unvisited sites
within range of previously visited sites.  Sweeping through the system
in a direction referred to as ``longitudinal,'' we only need keep
track of a ``transverse'' slice of the model.  This is illustrated in
Fig.~(\ref{latfig}).  Although the dimension of the Fock space is two
to the number of Grassmann variables, the storage requirements only
grow as two to the transverse volume.

The approach is exact, with no sign problems.  The complexity grows
severely with interaction range, probably limiting practical
applications to short range interactions in low dimensions.  Note that
the effort only grows linearly with the longitudinal dimension,
allowing very long systems.  This discussion has been in the context
of ``real'' Grassmann variables.  For ``complex'' variables treat
{$\psi$} and {$\psi^*$} independently.

In the transverse direction the boundary conditions are arbitrary, but
longitudinal boundaries should not be periodic.  To make them so
requires maintaining information on both the top and bottom layers of
the growing integration region, squaring the difficulty.  Note that
the technique is similar to the finite lattice method used for series
expansions \cite{series}, and closely related to a direct enumeration
of fermionic world lines \cite{worldline}.

As a test, consider a spin-less fermion hopping along a line of sites.
I introduce a complex Grassmann variable on each site of a two
dimensional lattice and study
\begin{equation}
Z=\int (d\psi d\psi^*) e^{S_t+S_h+S_I}
\end{equation}
with the various terms
\begin{equation}
\matrix{
S_t=\sum_{i,t} \psi^*_{i,t}(\psi_{i,t}-\psi_{i,t-1})\hfill \cr
S_h=k\sum_{i,t} \psi^*_{i,t}\psi_{i+1,t}+\psi^*_{i+1,t}\psi_{i,t}\hfill \cr
S_I=g\sum_{i,t} \psi^*_{i,t}\psi_{i,t}\psi^*_{i+1,t}\psi_{i+1,t}.\hfill\cr
}
\end{equation}
I take $N_t$ sites in the time $t$ direction and $N_i$ spatial sites.
The one-sided form of the temporal hopping insures an Hermitean
temporal transfer matrix\cite{transfer}.  This model Bosonizes into an
anisotropic quantum Heisenberg model, a fact not being used here.

\begin{figure}
\epsfxsize .92\hsize
\centerline {\epsfbox{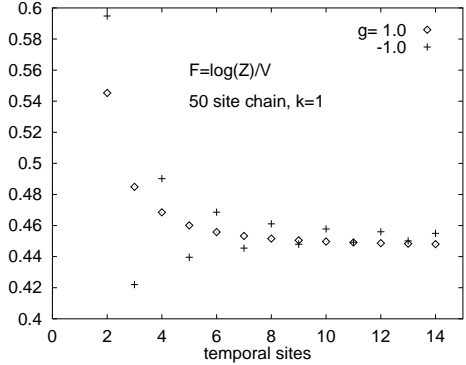}}
\vskip -.3in
\caption {The free energy $F=\log(Z)/ N_i N_t$ with a four fermion
interaction as described in the text, plotted as a function of the
number of time slices $N_t$.  The chain has $N_i=50$ sites.  Points
are shown for $k=1$ and $g=\pm 1$.}
\label{lambdafig}
\end{figure}

I treat $t$ as my ``transverse'' coordinate, growing the lattice along
the spatial chain.  Fig.~(\ref{lambdafig}) shows the $N_t$ dependence
for the free energy with $k=1$ and $g=\pm 1$.  Here memory
requirements were reduced by using time translation invariance after
integrating each layer.  The $N_t=14$ points were run on the RIKEN/BNL
Supercomputer.

\begin{figure}
\epsfxsize .92\hsize
\centerline {\epsfbox{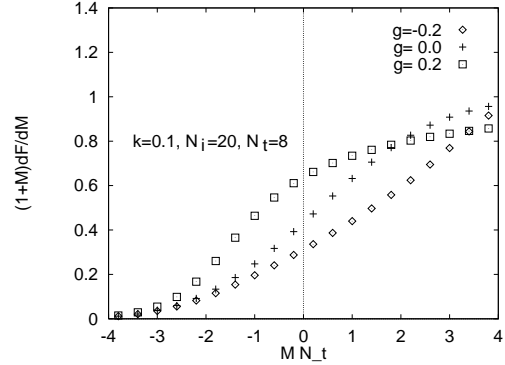}}
\vskip -.3in
\caption {The occupancy of a 20 site chain as a function of the
chemical potential scaled by the number of time slices.  The filling
occurs earlier or later depending on the sign of the coupling.}
\label{graph}
\end{figure}

With Monte Carlo methods, a chemical potential term can be highly
problematic due to cancelations.  Here, however, it is just another
local interaction of negligible cost.  As an illustration, take
$S=S_t+ S_h+ S_I+ S_M$ with
\begin{equation}
S_M=M\sum_{i,t} \psi^*_{i,t}\psi_{i,t}.
\end{equation}
This regulates the ``filling,'' which can be approximately monitored
as $(1+M) {dF\over dM}$.  I include the factor of $1+M$ to compensate
partially for finite $N_t$ artifacts.  Fig.~(\ref{graph}) shows the
filling as a function of $MN_t$ on an $N_t=8$ by $N_i=20$ lattice with
a spatial hopping parameter of $k=0.1$.  Here I made a crude
extrapolation in chain length by defining $F={1\over
N_t}\log\left({Z(N_i)\over Z(N_i-1)}\right)$ .  Note how the four
fermion coupling enhances the filling.

An obvious system for future study is the Hubbard model\cite{hubbard}.
This requires 4 Grassmann variables per site corresponding to $\psi^*$
and $\psi$ for spins up and down.  Higher spatial dimensions strongly
increase the size of the transverse volume and will limit practical
system volumes, but this may be compensated for by the lack of sign
problems.

\end{document}